\documentstyle[12pt]{article}
\author{W.~X.~Ma\thanks{On leave of absence from Institute of
Mathematics, Fudan University, Shanghai 200433, China}\,,
B.~Fuchssteiner and W.~Oevel \\
FB17, Mathematik-Informatik, Universitaet Paderborn,
\\ D-33098 Paderborn, Germany}
\title
{A $3\times 3$ matrix spectral problem for \\ AKNS hierarchy and
its binary Nonlinearization}
\date{\nonumber}

\begin{document}
\maketitle
%\Large

%\centerline{Abstract}

\begin{abstract} A three-by-three matrix spectral problem for AKNS
soliton hierarchy is proposed  and
the corresponding Bargmann symmetry constraint involved in Lax pairs and
adjoint Lax pairs is discussed.
The resulting nonlinearized Lax systems possess classical Hamiltonian
structures, in which the nonlinearized spatial system is intimately related
to stationary AKNS flows. These nonlinearized Lax systems also lead to a
sort of involutive solutions
to each AKNS soliton equation.
\end{abstract}

%%%%%%%%%%%%%%%%%%%%%%%%%%%%%%%%%%%%%%%%%%%%%%%%%%%%%%%%%%%%%%%%%%%%%%%%
\newcommand{\eqnsection}{
   \renewcommand{\theequation}{\thesection.\arabic{equation}}
   \makeatletter
   \csname $addtoreset\endcsname
   \makeatother}
\eqnsection

\newtheorem{thm}{Theorem}[section]
\newtheorem{Le}{Lemma}[section]
\newcommand{\R}{\mbox{\rm I \hspace{-0.9em} R}}

\def\be{\begin{equation}}
\def\ee{\end{equation}}
\def\ba{\begin{array}}
\def\ea{\end{array}}
\def\la {\lambda}
\def \part {\partial}
\def \al {\alpha}
\def \de {\delta}

%%%%%%%%%%%%%%%%%%%%%%%%%%%%%%%%%%%%%%%%%%%%%%%%%%%%%%%%%%%%%%%%%%%%%%%%

\section{Introduction}
\setcounter{equation}{0}

Symmetry constraints have aroused an increasing interest
 in recent few years due to the
important roles they play in soliton theory.
Such a kind of very successful symmetry
constraint method is the nonlinearization technique for Lax pairs of
soliton hierarchies,
including mono-nonlinearization proposed by Cao and Geng \cite{Cao}
\cite{CaoGeng} and further binary nonlinearization
\cite{MaStrampp} \cite{Ma} \cite{Geng}.

In general, one considers the complicated nonlinear problems to be solved
in such a way to break nonlinear problems into several linear or
smaller ones and then to solve these resulting problems.
It is following this idea that one has introduced the method of
 Lax pair to study nonlinear
soliton equations. The Lax pairs are always linear
with respect to their eigenfunctions. Nevertheless, the
 nonlinearization technique puts this original object, the Lax pair, into
a nonlinear and more complicated object, the nonlinearized Lax system.
It seems to be not reasonable enough, but in fact,
 it provides an effective way, different from the usual one, to solve soliton
equations. The main reason why the nonlinearization technique takes effect
is that kind of specific symmetry constraints expressed through the
variational derivative of the potential.

A similar symmetry constraint procedure for bi-Hamiltonian
soliton hierarchies
is presented by Antonowicz and Wojciechowski et al
\cite{AntonowiczWojciechowski1} \cite{AntonowiczWojciechowski2}
\cite{RagnicoWojciechowski}
and
bi-Hamiltonian
structures for the resulting classical integrable systems
can also be worked out through a Miura map \cite{AntonowiczWojciechowski2}
\cite{Blaszak}.
A connection between these systems and stationary flows \cite{BogoNovikov}
is also given by Tondo \cite{Tondo}
for the case of KdV hierarchy.
Because stationary
flows may be interpreted as finite dimensional Hamiltonian
systems \cite{BogoNovikov} based upon the so-called Jacobi-Ostrogradsky
coordinates \cite{Deleon},
a natural generalization of nonlinearization technique  to higher order
symmetry constraints is made by Zeng \cite{Zeng1} \cite{Zeng2} for the KdV
 and Kaup-Newell hierarchies etc.
 There have also been
some  algebraic geometric tricks,
proposed by Flaschka et al \cite{Flaschka} \cite{AdamsHP} \cite{Schilling},
to deal with similar
nonlinearized Lax pairs called Neumann systems.

The study of the nonlinearization theory leads to
a large class of interesting finite dimensional Liouville integrable
Hamiltonian systems which are connected with
soliton hierarchies (for example, see \cite{CaoGeng}
\cite{GengMa}). However in the literature, most results
are presented
for  the cases of $2\times 2$ matrix spectral problems.
The present
 paper is devoted to the symmetry constraints in binary nonlinearization
for
 a case of
$3\times 3$ matrix spectral problems.
We successfully propose a $3\times 3 $ matrix spectral problem for AKNS
soliton hierarchy, motivated by a representation of $3\times 3$ matrices
for the Lie algebra
$\textrm{sl}(2)$.
Then in Section 3, we consider the Bargmann symmetry constraint
 for the proposed new
Lax pairs and adjoint Lax pairs of AKNS soliton hierarchy.
In Section 4, we analyze the nonlinearized Lax systems, especially the
nonlinearized temporal systems, and establish a sort of
involutive  solutions to AKNS soliton equations.
Finally in Section 5, some remarks are given.

\section{New Lax pairs for AKNS equations}
\setcounter{equation}{0}

We introduce a  three-by-three matrix spectral problem
\be \phi_x=
\left(\begin{array}{c}
 \phi  _1 \\ \phi _2 \\ \phi _3\end{array}\right) _x=U(u,\la )
\left(\begin{array}{c}
 \phi _1 \\ \phi _2 \\ \phi _3\end{array}\right) =
\left(\begin{array}{ccc}
 -2\la & \sqrt{2}q&0 \vspace{1mm}\\ \sqrt{2}r&0 &\sqrt{2}q\vspace{1mm}
\\ 0&\sqrt{2}r &2\la
\end{array}\right)\left(\begin{array}{c}
 \phi _1 \\ \phi _2 \\ \phi _3\end{array}\right)
,\label{newsp}\ee
where the potential $u=(q  , r)^T$.
Its adjoint spectral problem reads as
\be \psi_x=
\left(\begin{array}{c}
 \psi  _1 \\ \psi _2 \\ \psi _3\end{array}\right) _x=-U^T(u,\la )
\left(\begin{array}{c}
 \psi _1 \\ \psi _2 \\ \psi _3\end{array}\right) =
\left(\begin{array}{ccc}
 2\la &- \sqrt{2}r&0 \vspace{1mm}\\ -\sqrt{2}q&0 &-\sqrt{2}r\vspace{1mm}
\\ 0&-\sqrt{2}q &-2\la
\end{array}\right)\left(\begin{array}{c}
 \psi _1 \\ \psi _2 \\ \psi _3\end{array}\right)
.\label{newasp}\ee
Here $T$ means the transposition of the matrix.
Our purpose is to generate AKNS hierarchy of soliton equations from
the above specific spectral problem (\ref{newsp}).
To this end, we first solve the adjoint representation equation $V_x=[U,V]$.
Take
\be
V=
\left(\begin{array}{ccc} 2a&\sqrt{2}b&0 \vspace{1mm}\\ \sqrt{2}c&0&\sqrt{2}b
\vspace{1mm}\\
0&\sqrt{2}c&-2a\end{array}\right)
=\sum _{i=0}^\infty
\left(\begin{array}{ccc} 2a_i&\sqrt{2}b_i&0\\\sqrt{2}c_i&0&\sqrt{2}b_i\\
0&\sqrt{2}c_i&-2a_i\end{array}\right) \la ^{-i}
\ee
and then we have
\[
[U,V]=\left(\begin{array}{ccc} 2(qc-rb)&-2\sqrt{2}(\la b+qa)&
0 \vspace{1mm}\\ 2\sqrt{2}(ra+\la c)&0&-2\sqrt{2}(\la b+qa)
\vspace{1mm}\\ 0&2\sqrt{2}(ra+\la c)&-2(qc-rb)
\end{array}\right) .\]
Therefore we easily find
that the adjoint representation equation $V_x=[U,V]$ becomes
\[a_x=qc-rb,\ b_x=-2\la b-2qa,\ c_x=2\la c+2ra,\]
which is equivalent to
\be
a_{ix}=q c_i-r b_i,\
 b_{ix}=-2b_{i+1}-2q a_i,\
c_{ix}=2c_{i+1}+2r a_i, \ i\ge0.\label{recursionre0}\ee
We fix the initial values
 \be a_0=-1,\, b_0=c_0=0\ee
 and require that
\be a_i|_{u=0}=
b_i|_{u=0}=
c_i|_{u=0}=0,\ i\ge 1,\label{zerocondition}\ee
which equivalently select constants of integration to
be zero. On the other hand, the above equality
(\ref{recursionre0}) gives rise to
 the recursion relation for determining $a_i,b_i,c_i$:
\be \left\{ \begin{array} {l}
a_{i+1}=\frac12 \part ^{-1}(qc_{ix}+rb_{ix}),\\
b_{i+1}=-\frac12 b_{ix}-qa_i,\\
c_{i+1}=\frac12 c_{ix}-ra_i,
\end{array}\right.\ \, i\ge0.
\label{recursionre}\ee
This recursion relation
uniquely determines infinitely many  sets of polynomials
$a_i,b_i,c_i,\ i\ge 1$,
in $u,u_x,\cdots$ under the requirement
(\ref{zerocondition}).
The first two sets are as follows
\[a_1=0, \ b_1=q ,\ c_1=r ;\
a_2=\frac12 (qr ),\
b_2=-\frac12 q _x,\
c_2=\frac12 r _x.\]
In addition, we have
\[a^2+bc=(\sum_{i=0}^\infty a_i\la ^{-i})^2+(\sum_{i=0}^\infty b_i\la ^{-i})
(\sum_{i=0}^\infty c_i\la ^{-i})=1,
\]
because
$(a^2+bc)_x=\frac18\textrm{tr}(V^2)_x=\frac18 \textrm{tr}[U,V^2]=0$
and $(a^2+bc)|_{u=0}=1.$ It follows that $a_i,b_i,c_i,\ i\ge 1$, are local.

A direct computation  may show that the compatibility
conditions of the Lax pairs
\be \phi _x=U\phi ,\ \phi _{t_n}=V^{(n)}\phi ,\ V^{(n)}=V^{(n)}(u,\la )
=(\la ^nV)_+,\ n\ge
0,\label{laxpair}\ee
or the adjoint Lax pairs
\be \psi _x=-U^T\phi ,\ \psi _{t_n}=-(V^{(n)})^T
\psi ,\ n\ge
0,\label{alaxpair}\ee
where the symbol $+$ denotes the choice of non-negative power of
$\la $, engenders a hierarchy of AKNS soliton equations
\be u_{t_n}= \left(\ba {c}
 q \\r \ea \right)
_{t_n}=K_n=
\left(\ba {c}
-2b_{n+1}\\2c_{n+1}
\ea  \right)
=JL^n \left(\ba {c} r\\q \ea \right)
,\ n\ge 0,\label{akns}\ee
where the Hamiltonian operator $J$ and the recursion operator $L$
read as
\be J= \left(\ba {cc} 0&-2\\2&0\ea \right) ,\
L= \left(\ba {cc} \frac12 \part -r \part ^{-1}q &r \part ^{-1}r \\
-q \part ^{-1}q &-\frac12 \part +q \part ^{-1}r \ea \right) .\ee
This AKNS hierarchy is exactly the same as one in Ref. \cite{MaStrampp},
which also shows that the same soliton hierarchy may possess different
Lax pairs, even different order spectral matrices.
Here the operator $L^*$ is a hereditary operator \cite{FuchssteinerFokas},
and $J$ and $JL$ constitute a Hamiltonian pair.

Finally, we would like to elucidate the other two properties
on AKNS hierarchy (\ref{akns}). First by Corollary 2.1 of Ref.
\cite{MaStrampp},
we can obtain
\be V_{t_n}=[V^{(n)},V],\ n\ge 0\ee
when $u_{t_n}=K_n$, i.e. $U_{t_n}-V_x^{(n)}+[U,V^{(n)}]=0,\ n\ge0$.
Second, we can get the Hamiltonian structure of AKNS hierarchy
\be u_{t_n}=
K_n=J
\left(\begin{array}{c} c_{n+1}\\b_{n+1}\end{array}\right)
=J\frac {\delta H_n}{\delta u}
,\  H_n=\frac 2{n+1}a_{n+2},\ n\ge 0,
\ee
by applying the trace identity \cite{Tu} \cite{TuAH}.

\section{Binary nonlinearization related to new spectral problem}
\setcounter{equation}{0}

In order to impose the Bargmann symmetry constraint
in binary nonlinearization,
we first need to compute the variational derivative of the
spectral parameter $\la $
with to the potential $u$, which is shown in the following Lemma
\cite{FokasAnderson}\cite{MaStrampp}.

\begin{Le}
Let $U(u,\la )$ be a matrix of order $s$ depending on $u,\, u_x,\,
\cdots$ and a parameter $\la $. Suppose that
$\phi =(\phi _1,\phi_2,\cdots, \phi_s)^T,$ $\psi =(\psi _1,\psi_2,
\cdots, \psi_s)^T$ satisfy the spectral problem and the adjoint spectral
problem
\[ \phi _x=U(u,\la )\phi,\ \psi _x=-U^T(u,\la )\psi,\]
and set the matrix
$\bar V =\phi \psi ^T= (\phi _k\psi _l)_{s\times s }$,
then we have the following two results:

\noindent (i) the variational derivative of the spectral parameter $\la $
with respect to the potential $u$ may be expressed by
\be
 \frac {\delta \la }{\delta u}=
\frac {\textrm{tr}\bigl(\bar V\frac{\part U}{\part u}\bigr)}{
-\int _{-\infty}^\infty \textrm{tr}\bigl(\bar V\frac {\part U}{\part \la
}\bigr) dx},\label{vdoflambda}
\ee

\noindent (ii) the matrix
$\bar V $
is a solution to the adjoint representation equation
 $ V _x=[U, V]$, i.e. $\bar V _x=[U,\bar V]$.
\label{FAMS}\end{Le}

Following (\ref{vdoflambda}), we have the
variational derivative of the spectral parameter for
the spectral problem
(\ref{newsp}) and the adjoint spectral problem (\ref{newasp})
\be
\frac {\delta \la }{\de q}=\frac{\sqrt{2}}E(\phi _2\psi _1+\phi _3\psi_2),\
\frac {\delta \la }{\de r}=\frac{\sqrt{2}}E(\phi _1\psi _2+\phi _2\psi_3),
\label{dvakns}\ee
where $E=2\int_{-\infty}^\infty (\phi _1\psi _1-\phi _3\psi_3)dx$.

Let us  introduce $N$ $(N\ge 1)$ distinct eigenvalues
$\la _j,\, 1\le j\le N$, and denote by
\[ \phi^{(j)}=(\phi_{1j},\phi_{2j},\phi_{3j})^T, \
\psi^{(j)}=(\psi_{1j},\psi_{2j},\psi_{3j})^T,\ 1\le j\le N,
\]
the eigenfunctions of (\ref{laxpair}) and the adjoint
eigenfunctions of (\ref{alaxpair}), i.e.
\be
\phi_x^{(j)}=U(u,\la _j)\phi ^{(j)},\ \psi_x^{(j)}=-U^T(u,\la _j)\psi ^{(j)},
\ 1\le j\le N,\label{xpart}
\ee
\be
\phi_{t_n}^{(j)}=V^{(n)}(u,\la _j)\phi ^{(j)},\ \psi_{t_n}^{(j)}=
-(V^{(n)})^T(u,\la _j)\psi ^{(j)},
\ 1\le j\le N.\label{tpartn}
\ee
 Now we make
the Bargmann symmetry constraint
\be
K_0=J\frac {\delta H_0 }{\delta u}
=J\sum _{j=1}^N \mu _jE_j \frac {\delta \la _j}{\delta u}
,\label{Bargmannsym}\ee
where $ E_j=2\int ^\infty_{-\infty}(\phi  _{1j}\psi_{1j}
 -\phi _{3j}\psi_{3j})dx,
\ 1\le j\le N,$ and $\mu _j,\ 1\le j\le N,$ are any nonzero constants.
By (\ref{dvakns}), this symmetry constraint becomes
\[
K_0=J\sum
_{j=1}^N \mu _j\left(\begin{array}{c} \sqrt{2}(\phi _{2j}\psi _{1j}+
\phi_{3j}\psi_{2j})\vspace{1mm}\\ \sqrt{2}(\phi _{1j}\psi_{2j}+
\phi_{2j}\psi_{3j})\end{array}\right),\]
from which
we get the following
explicit expression for the potential $u$
\be u
=f(P_1,P_2,P_3;Q_1,Q_2,Q_3)=\sqrt{2}\left(\begin{array}{c}
< P  _1,BQ_2>+< P  _2,BQ_3>\\
< P  _2,BQ  _1>+< P  _3,BQ  _2>
\end{array}\right).\label{expressionofu}\ee
Here and hereafter,
 $<\cdot,\cdot>$ denotes the standard inner product of $\R ^N$ and
\be B=\textrm{diag}(\mu _1,\cdots, \mu _N),\
 \left(\begin{array}{c} P _i\\  Q _i\end{array}\right)=
\left(\begin{array}{c} (\phi _{i1},\phi _{i2},\cdots,\phi _{iN})^T
\vspace {1mm}\\
(\psi_{i1},\psi_{i2},\cdots,\psi_{iN})^T\end{array}\right)
,\  i=1,2,3 \label{PQB}
.\ee
The substitution of  (\ref{expressionofu}) into the spatial system
(\ref{xpart}) and the
temporal systems (\ref{tpartn}) for $n\ge 0$  yields the
 nonlinearized spatial system:
\be \left\{\begin{array}{l}
\left(\begin{array}{c}
\phi _{1j}\\ \phi _{2j}\\ \phi_{3j}\end{array}\right)
_x=U(f,\la _j)
\left(\begin{array}{c} \phi _{1j}\\ \phi _{2j}
\\ \phi_{3j}\end{array}\right) ,\ j=1,2,\cdots,N,\vspace {3mm}\\
\left(\begin{array}{c} \psi_{1j}\\ \psi_{2j}\\ \psi_{3j}\end{array}\right)
_x=-U^T(f,\la _j)
\left(\begin{array}{c} \psi_{1j}\\ \psi_{2j}\\ \psi_{3j}\end{array}\right) ,\
j=1,2,\cdots,N;
\end{array}
\right.\label{nxpart}\ee
and the nonlinearized temporal systems for $n\ge 0$:
\be \left\{\begin{array}{l}
\left(\begin{array}{c} \phi _{1j}\\ \phi _{2j}\\ \phi_{3j}\end{array}\right)
_{t_n}=V^{(n)}(f,\la _j)
\left(\begin{array}{c} \phi _{1j}\\ \phi _{2j}\\ \phi_{3j}\end{array}\right) ,\
j=1,2,\cdots,N,\vspace {3mm}\\
\left(\begin{array}{c} \psi_{1j}\\ \psi_{2j}\\ \psi_{3j}\end{array}\right)
_{t_n}=-(V^{(n)})^T(f,\la _j)
\left(\begin{array}{c} \psi_{1j}\\ \psi_{2j}\\ \psi_{3j}\end{array}\right) ,\
j=1,2,\cdots,N.
\end{array}
\right.\label{ntpartn}\ee
It is obvious that
(\ref{nxpart}) is a system of ordinary differential equations and
(\ref{ntpartn})
is a hierarchy of partial differential equations.

Suppose that $Z $ is an expression depending on $u$ and its differentials.
{}From now on we use $\widetilde Z $ to denote the expression of $Z $
depending
on $P_i,Q_i,\ 1\le i\le 3,$ and their differentials
 after substituting
(\ref{expressionofu}) into $Z $, and use $\textrm{Or}(\widetilde Z )$ to denote
the expression of $\widetilde Z $ only depending
on $P_i,Q_i,\ 1\le i\le 3,$ themselves after substituting
(\ref{nxpart}) into $\widetilde Z $ sufficiently many times.
Therefore  (\ref{ntpartn})
may be transformed into the following systems for $n\ge 0$:
 \be \left\{\begin{array}{l}
\left(\begin{array}{c} \phi _{1j}\\ \phi _{2j}\\ \phi_{3j}\end{array}\right)
_{t_n}=\textrm{Or}( V^{(n)}(f,\la _j))
\left(\begin{array}{c} \phi _{1j}\\ \phi _{2j}\\ \phi_{3j}\end{array}\right)
 ,\ j=1,2,\cdots,N,\vspace {3mm}\\
\left(\begin{array}{c} \psi_{1j}\\ \psi_{2j}\\ \psi_{3j}\end{array}\right)
_{t_n}=-(\textrm{Or}( V^{(n)}(f,\la _j))^T
\left(\begin{array}{c} \psi_{1j}\\ \psi_{2j}\\ \psi_{3j}\end{array}\right)
 ,\ j=1,2,\cdots,N,
\end{array}
\right.\label{nxtpartn}\ee
which are all ordinary differential equations with an independent
variable $t_n$ because the matrices
$\textrm{Or}( V^{(n)}(f,\la _j)),\ n\ge 0,\ 1\le j\le N,$
 only depend on $P_i,Q_i,\  1\le i\le 3.$

We would like to discuss the integrability
on the nonlinearized spatial system (\ref{nxpart})
and the nonlinearized temporal systems (\ref{nxtpartn}) for $n\ge 0$
in the Liouville sense \cite{Perelomov}.
We shall utilize the
symplectic structure $\omega ^2$  on $\R ^{6N}$
\[ \omega ^2=\sum _{i=0}^3\sum _{j=0}^N \mu _j d\phi_{ij}\wedge d\psi_{ij}=
\sum _{i=0}^3(BdP_i)\wedge dQ_i,\]
by which one can define the corresponding Poisson bracket
for two functions $F,G$ defined over the phase space  $\R ^{6N}$
\begin{eqnarray}\{F,G\}&=&\omega ^2(IdG,IdF)=\omega ^2(X_G,X_F)\nonumber\\
&=&
\sum_{i=1}^3 \sum _{j=1}^N\mu _j^{-1}
(\frac {\part F}{\part \psi_{ij}}
\frac {\part G}{\part \phi _{ij}}-
\frac {\part F}{\part \phi _{ij}}
\frac {\part G}{\part \psi_{ij}})\nonumber\\&=&
\sum_{i=1}^3(<\frac {\part F}{\part Q_i},B^{-1}\frac {\part G}{\part P_i}>-
<\frac {\part F}{\part P_i},B^{-1}\frac {\part G}{\part Q_i}>),
\label{poissonbracket}
\end{eqnarray}
where $IdH=X_H$ denotes the Hamiltonian vector field
 with energy $H$ determined by
\[\omega ^2(X ,IdH)=\omega ^2(X, X_H)=dH(X ),\  X \in T(\R ^{6N}),\]
 and
the corresponding Hamiltonian system with the Hamiltonian function $H$
\be \dot {x}=IdH(x)=dx(IdH)=\omega ^2\{IdH, Idx\}=\{x,H\},\ x\in \R ^{6N},\ee
which possesses an explicit formulation
\be
\dot {P}_i=-B^{-1}\frac {\part H}{\part Q_i},\
\dot {Q}_i=B^{-1}\frac {\part H}{\part P_i},\ i=1,2, 3.\ee
Note that there are some authors who use the other Poisson bracket $\{F,G\}=
\omega ^2(X_F,X_G)$. As remarked by Carroll \cite{Carroll},
it doesn't matter of course but each type has many proponents and
hence one must be careful of minus signs in reading various sources.
The notation we accept here is the Arnold's one \cite{Arnold}.

\begin{thm}
The following functions
\be \bar F_j=\sum_{i=1}^3\phi_{ij}\psi_{ij},\
1\le j\le N,\ee
are all integrals of motion for
the nonlinearized spatial system (\ref{nxpart}). Moreover they are
in involution under the Poisson bracket (\ref{poissonbracket})
and independent over
the region
\[\Omega = \{ \R^{6N}| \,\phi_{ij},\psi_{ij}\in \R,\
 \sum_{i=1}^3(\phi_{ij}^2+\psi_{ij}^2)
\ne 0,\ 1\le j\le N\}.\]
\end{thm}
{\bf Proof}: Let
\[
\bar V(\la _j)=
(\phi_{kj}\psi_{lj})_{k,l=1,2,3}\,,\ 1\le j\le N.\]
We can first find that
\[ \bar F_j =  \textrm{tr} \bigl(\bar V(\la _j)\bigr).\]
On the other hand, by Lemma
\ref{FAMS}
we know that $\bar V(\la _j)$ satisfies
\[\bar V(\la _j)_x=[U(\widetilde u,\la _j),\bar V(\la _j)]\] when
(\ref{nxpart}) holds, and thus
\[ \ba {l}\bar F_{jx}=( \textrm{tr} (\bar V(\la _j)))_x=
\textrm{tr} (\bar V(\la _j))_x\\
=
\textrm{tr}[U(\widetilde u,\la _j),\bar V(\la _j)]=0,\ea \]
which shows that $\bar F_{j},\, 1\le j\le N$,
are all  integrals of motion  for the nonlinearized spatial system
(\ref{nxpart}).
In addition, it is very easy to prove that
\[\{\bar F_k,\bar F_l\}=0,\ 1\le k,l\le N,\]
which means $\bar F_j,\ 1\le j\le N,$ are in involution.
It is also obvious that $\textrm{grad}\bar F_j, \, 1\le j\le N,$ are
everywhere
linear
independent over $\Omega$ by observing that
\begin{eqnarray}
\left(\ba {c}
\frac {\part \bar F_k}{\part \phi_{il}}
\ea \right)_{k,l=1,\cdots,N}&=&\left(\ba {cccc}
\psi_{i1}&  &  &0\\
 & \psi_{i2}& & \\
 & & \ddots & \\
0 & & & \psi_{iN}\ea \right),\ i=1,2,3,\nonumber\\
\left(\ba {c}
\frac {\part \bar F_k}{\part \psi_{il}}
\ea \right)_{k,l=1,\cdots,N}&=&\left(\ba {cccc}
\phi_{i1}&  &  &0\\
 & \phi_{i2}& & \\
 & & \ddots & \\
0 & & & \phi_{iN}\ea \right),\ i=1,2,3.\nonumber\end{eqnarray}
The proof is completed.
$\vrule width 1mm height 3mm depth 0mm$

Throughout our paper, we assume that
\be A=\textrm{diag}(\la _1,\la _2,\cdots,\la _N).\ee
If all elements $Z_{ij},\ 1\le i,j\le s$
of a given matrix $Z=(Z_{ij})_{s\times s }$
are  polynomials in $\la $, i. e. $Z_{ij}=\sum _{k=0}^mZ_{ijk}\la ^k$,
for convenience of presentation,
we define a new matrix called $M_A(Z)$ as follows
\be
M_A(Z)=(
\sum _{k=0}^mZ_{ijk}A ^k)_{sN\times sN}.\label{M}\ee
Moreover we often accept compact forms, for example,
\[ \frac \part {\part P_i}=(\frac \part {\part \phi_{i1}},\cdots,
\frac \part {\part \phi_{iN}})^T,\
\{P_i,H\}=(\{\phi_{i1},H\},\cdots,\{\phi_{iN},H\})^T,
\  i=1,2, 3.\]

\begin{thm} We have the explicit integrals
of motion for the nonlinearized spatial system (\ref{nxpart}):
 \be \left \{\begin{array}{l}
F_1=-8(<P_1 ,BQ_1>-<P_3,BQ_3>),\vspace{2mm}\\
F_m=4\sum _{i=1}^{m-1}\bigl[
(<A^{i-1}P _1,BQ _1>-<A^{i-1}P_3,BQ_3>)\times \vspace{2mm}\\
\qquad (<A^{m-i-1}P _1,BQ _1>-<A^{m-i-1}P_3,BQ_3>)\vspace{2mm}
\\ \qquad
+ 2(<A^{i-1}P_1,BQ_2>+<A^{i-1}P_2,BQ_3>)\times \vspace{2mm}\\
\qquad (<A^{m-i-1}P_2,BQ_1>+<A^{m-i-1}
P_3,BQ_2>)\bigr]\vspace{2mm}\\ \qquad
-8(<A^{m-1}P _1,BQ _1>-<A^{m-1}P_3,BQ_3>),
\ m\ge 2,\end{array}\right.\label{F_m}\ee
where $P_i,\, Q_i,\, B$ are defined by (\ref{PQB}).
Moreover they constitute an involutive system together with
$\bar F_j,\ 1\le j\le N$, under the Poisson bracket (\ref{poissonbracket}),
 i.e.
\[\{F_{k},F_{l}\}=\{F_{m},\bar F_{j}\}=0, \ m,
k,l\ge 1,\ 1\le j\le N.\]
\end{thm}
{\bf Proof}:
We assume that
\begin{eqnarray}
&&
 \hat {q}
=\sqrt{2}(
< P  _1,BQ_2>+< P  _2,BQ_3>),\  \hat {r}=\sqrt{2}(
< P  _2,BQ  _1>+< P  _3,BQ  _2>)
;\nonumber\\
&&\hat {a}_0=-1,\ \hat {b}_0=\hat {c}_0=0;
\nonumber\\
&&\left\{\ba {l}\hat {a}_{i+1}=<A^{i}P_1,BQ_1>-<A^{i}P_3,BQ_3>, \ i\ge 0,
\vspace {1mm}\\
\hat {b}_{i+1}=\sqrt{2}(<A^{i}P_1,BQ_2>+<A^{i}P_2,BQ_3>), \ i\ge 0,
\vspace {1mm}\\
\hat {c}_{i+1}=\sqrt{2}(<A^{i}P_2,BQ_1>+<A^{i}P_3,BQ_2>), \ i\ge 0.
\ea \right. \label{hatabci}\end{eqnarray}
Further we choose
that
\[\hat {U}=\left (\ba {ccc}
-2\la  &\sqrt{2}\hat{q}&0\\ \sqrt{2}\hat{r}&0&\sqrt{2}\hat{q}\\
0&\sqrt{2}\hat{r}&2\la
\ea \right),\
\hat {V}=\left (\ba {ccc}
2\hat{a} &\sqrt{2}\hat{b}&0\\ \sqrt{2}\hat{c}&0&\sqrt{2}\hat{b}\\
0&\sqrt{2}\hat{c}&-2\hat{a}
\ea \right) ,\]
where $\hat{a},\  \hat{b}$ and $  \hat{c}$ are defined by
\[
\hat{a}=\sum_{i=0}^\infty \hat{a}_i \la ^{-i},\
\hat{b}=\sum_{i=0}^\infty \hat{b}_i\la ^{-i}
,\ \hat{c}=\sum_{i=0}^\infty \hat{c}_i \la ^{-i}.\]
It may be shown that when the nonlinearized spatial system
(\ref{nxpart}) holds, we have
\[ \hat{V}_x=[\hat{U},\hat{V}],\  \textrm{i.e.}\
\hat{a}_x=\hat{q}\hat{c}-\hat{r}\hat{b},\ \hat{b}_x=-2\la \hat{b}
-2\hat{q}\hat{a},\ \hat{c}_x=2\la \hat{c}+2\hat{r}\hat{a}.
\]
Therefore we can compute that
\[
\hat {F}_x:=(\frac12 \textrm{tr}(\hat{V}^2))_x=\frac12 \textrm{tr}
(\hat{V}^2)_x=\frac12 \textrm{tr}[
\hat{U},\hat{V}^2]=0.\]
On the other hand, we have
\[ \hat {F}=4(\hat{a}^2+\hat{b}\hat{c})=
 \sum_{m=0}^\infty F_m \la ^{-m},\ F_0=4,\
F_m=4\sum _{i=0}^m(\hat{a}_i\hat{a}_{m-i}+\hat{b}_i\hat{c}_{m-i}),\ m\ge1.
\]
Hence $F_m,\ m\ge1$, are all integrals of motion for
the nonlinearized spatial system
(\ref{nxpart}).

Now we turn to the involutivity of integrals of motion.
We take
\[
\hat {V}^{(n)}(\la )=(\la ^n\hat {V})_+\]
and construct a temporal system for $n\ge0$
\be
\left(\ba {c} P_1\\ P_2\\ P_3\ea \right)_{t_n}=M_A(\hat {V}^{(n)})
\left(\ba {c} P_1\\ P_2\\ P_3\ea \right),\
\left(\ba {c} Q_1\\ Q_2\\ Q_3\ea \right)_{t_n}=-(M_A(\hat {V}^{(n)}))^T
\left(\ba {c} Q_1\\ Q_2\\ Q_3\ea \right),\label{hatntpartn}\ee
where $M_A(\hat{V}^{(n)})$, $n\ge0$, are determined in the way of (\ref{M}).
We can first prove that
when this system (\ref{hatntpartn}) holds, we have
\[
(\hat {V}(\la ))_{t_n}=[\hat {V}^{(n)}(\la ),\hat {V}(\la )].\]
Therefore $F_m,\ m\ge1$, are also integrals of motion for
the system (\ref{hatntpartn}). Secondly,
we can verify that
\[
\left(\ba {c}
B^{-1}\frac {\part \hat{F}}{\part P_1}\vspace{1mm}\\
B^{-1}\frac {\part \hat{F}}{\part P_2}\vspace{1mm}\\
B^{-1}\frac {\part \hat{F}}{\part P_3}\ea \right)=
\left(\ba {c}
B^{-1}\textrm{tr}(\hat{V}\frac \part {\part P_1}\hat{V})\vspace{1mm}\\
B^{-1}\textrm{tr}(\hat{V}\frac \part {\part P_2}\hat{V})\vspace{1mm}\\
B^{-1}\textrm{tr}(\hat{V}\frac \part {\part P_3}\hat{V})\ea \right)=
\sum_{m=0}^\infty (M_A(\hat {V}^{(m)}))^T
\left(\ba {c} Q_1\vspace{1mm}\\ Q_2\vspace{1mm}\\
Q_3\ea \right) \la ^{-m-1},\]
\[
\left(\ba {c}
B^{-1}\frac {\part \hat{F}}{\part Q_1}\vspace{1mm}\\
B^{-1}\frac {\part \hat{F}}{\part Q_2}\vspace{1mm}\\
B^{-1}\frac {\part \hat{F}}{\part Q_3}\ea \right)=
\left(\ba {c}
B^{-1}\textrm{tr}(\hat{V}\frac \part {\part Q_1}\hat{V})\vspace{1mm}\\
B^{-1}\textrm{tr}(\hat{V}\frac \part {\part Q_2}\hat{V})\vspace{1mm}\\
B^{-1}\textrm{tr}(\hat{V}\frac \part {\part Q_3}\hat{V})\ea \right)=
\sum_{m=0}^\infty (M_A(\hat {V}^{(m)}))
\left(\ba {c} P_1\vspace{1mm}\\ P_2\vspace{1mm}\\
P_3\ea \right) \la ^{-m-1}.\]
These two equalities show that the system (\ref{hatntpartn}) for $n\ge 0$
are all Hamiltonian systems with Hamiltonian functions $-F_{n+1}$.
Therefore
\[\{F_{m+1},-F_{n+1}\}=\frac d {d t_n}F_{m+1}=0,\ m,n\ge0,\]
which shows the involutivity of $F_m,\,m\ge1$.
In addition, it is easy to get that
\[ \{F_k,\bar F_l\}=\sum_{i=0}^3\mu _l^{-1}
(\frac {\part F_k}{\part \psi_{il}}\psi_{il}
-\frac {\part F_k}{\part \phi_{il}}\phi_{il})=0,\ k\ge 1,\ 1\le l\le N,\]
noting the particular form of $F_k,\,k\ge1$.
The proof is finished.
 $\vrule width 1mm height 3mm depth 0mm$

Because we have
 $\bar{V}^2(\la _j)=
\textrm{tr}(\bar{V}(\la _j))\bar V(\la _j),$ $1\le j\le N$,
$\hat {V}^3=\frac12\textrm{tr}(\hat {V}^2)\hat {V}$, we cannot obtain new
integrals of motion of the nonlinearized spatial system (\ref{nxpart})
from the trace of other power of $\bar V(\la _j)$ and $\hat {V}$.
In addition to this, it is interesting to observe that the determinants of
the matrices $\bar V(\la _j),$ $1\le j\le N$, and $\hat {V}$ are all zero.

The nonlinearized
spatial system is easily rewritten as an  Hamiltonian system
\[ P  _{ix}=\{P_i,H\}=-B^{-1}\frac {\part H}{\part  Q  _{i}},\
 Q  _{ix}=\{Q_i,H\}=B^{-1}\frac {\part H}{\part  P  _i},\
i=1,2,3\]
with the Hamiltonian function
\[\begin{array}{l}H=2(<A P _1,BQ  _1>-<AP _3,BQ  _3>)\vspace{2mm}\\
-2(<P _1,BQ  _2>+<P _2,BQ  _3>)(<P _2,BQ  _1>+<P _3,BQ  _2>)
=-\frac14 F_2+\frac
1 {64}F_1^2
.\end{array}\]
Thus it  possesses
the following $3N$ involutive integrals of motion
\[ \bar F_j,\ 1\le j\le N, F_m,\ 1\le m\le 2N.\]
In some special cases, they
may be shown to be independent at least on certain region.

\begin{thm}
When $N=1,2$, the determinant of the matrix
\[D(N)=
\left(\ba {ccc}
(\textrm{grad}_{P_1}\bar F_1)^T&(\textrm{grad}_{P_2}\bar F_1)^T&
(\textrm{grad}_{P_3}\bar F_1)^T\\
\vdots &\vdots &\vdots \\
(\textrm{grad}_{P_1}\bar F_N)^T&(\textrm{grad}_{P_2}\bar F_N)^T&
(\textrm{grad}_{P_3}\bar F_N)^T\\
(\textrm{grad}_{P_1} F_1)^T&(\textrm{grad}_{P_2} F_1)^T&
(\textrm{grad}_{P_3} F_1)^T\\ \vdots &\vdots &\vdots \\
(\textrm{grad}_{P_1} F_{2N})^T&(\textrm{grad}_{P_2} F_{2N})^T&
(\textrm{grad}_{P_3} F_{2N})^T
\ea \right)\]
with $\textrm{grad}_{P_i} G=(\frac {\part G}{\part \phi_{i1}},\cdots,
\frac {\part G}{\part \phi_{iN}})^T,\ 1\le i\le 3,$
is not always zero. Thus
the integrals of motion
$\bar F_j,\ 1\le j\le N, F_m,\ 1\le m\le 2N,$ are
independent at least on some region.
\end{thm}
{\bf Proof}: We have
\begin{eqnarray} D(1)&=&
(16 \phi_{11} \psi_{11} \psi_{21}^3  + 16\phi_{31} \psi_{21}^3  \psi_{31} -
 32 \phi_{11} \psi_{11 }^2
\psi_{21} \psi_{31}\nonumber
   \\ &&
+ 32 \phi_{21} \psi_{11} \psi_{21 }^2 \psi_{31} - 32 \phi_{31} \psi_{11}
 \psi_{21 }\psi_{31 }^2
   - 64 \phi_{21} \psi_{11}^2  \psi_{31 }^2)\mu _1^3\nonumber\end{eqnarray}
and \[
 D(2)=
         (-256 \la _1^3  + 256 \la _2^3  - 768 \la _1 \la _2 ^2 + 768 \la _1
  ^2\la _2)\mu _1^3\mu _2^4,\]
when we choose
\[
\ba {l}
\phi_{11}=1,\  \phi_{12}=0,\
 \phi_{21}=0,\ \phi_{22}=-1,\
 \phi_{31}=0,\ \phi_{32}=0,\vspace{1mm}\\
 \psi_{11}=0,\ \psi_{12}=-1,\
 \psi_{21}=1,\ \psi_{22}=0,\
  \psi_{31}=0,\ \psi_{32}=-1.\ea \]
These are consequences of computation by the computer algebra system
MuPAD. But they may also be
shown by some  direct computation.
The proof is completed.
$\vrule width 1mm height 3mm depth 0mm$

According to the above theorem,
the nonlinearized
spatial system (\ref{nxpart}) is Liouville integrable on some
region of the phase space, when $N=1,2$.

\section{
Involutive solutions}
\setcounter{equation}{0}

The aim of this section is to discuss some properties on the nonlinearized
spatial system (\ref{nxpart}) and the nonlinearized
spatial system (\ref{nxtpartn}) and to establish a kind of
involutive solutions with separated variables for AKNS soliton equations.
\begin{Le}\label{lemmarealre}
When $\phi=(\phi_1,\phi_2,\phi_3)^T$ and $\psi=(\psi_1,\psi_2,\psi_3)^T $
satisfy the spectral problem (\ref{newsp}) and the adjoint spectral problem
(\ref{newasp}), we have
\be
L\left ( \ba {c} \phi_2\psi_1+\phi_3\psi_2\\
\phi_1\psi_2+\phi_2\psi_3\ea \right)=\la
\left ( \ba {c} \phi_2\psi_1+\phi_3\psi_2\\
\phi_1\psi_2+\phi_2\psi_3\ea \right)+I\left ( \ba {c}
r\\q
\ea \right),\label{realre}\ee
where $I$ is an integral of motion for (\ref{newsp}) and (\ref{newasp}).
\end{Le}
{\bf Proof}:
{}From the spectral problem (\ref{newsp}) and the adjoint spectral problem
(\ref{newasp}), we can find that
\[
\part (\phi_1\psi_1-\phi_3\psi_3)=\sqrt{2} q (\phi_2\psi_1+\phi_3\psi_2)
-\sqrt{2}
r (\phi_1\psi_2+\phi_2\psi_3).\]
This  yields
\[\part ^{-1}[-q (\phi_2\psi_1+\phi_3\psi_2)
+
r (\phi_1\psi_2+\phi_2\psi_3)]=-\frac1 {\sqrt{2}}(\phi_1\psi_1-\phi_3\psi_3)
+I,\]
where $I$ is an integrals of motion for (\ref{newsp}) and (\ref{newasp}).
The relation (\ref{realre}) follows from the above equality.
$\vrule width 1mm height 3mm depth 0mm$

We recall that $\widetilde Z$ denotes the expression of $Z$ depending on $P_i,
Q_i,\,1\le i\le 3,$ and their differentials
after the substitution of (\ref{expressionofu}) and
 into $Z$,
and that $\textrm{Or}(\widetilde Z)$ denotes
the expression of $\widetilde Z$ only depending
on $P_i,Q_i,\ 1\le i\le 3,$ themselves after the substitution of
(\ref{nxpart}) into $\widetilde Z$ sufficiently many times.
A general result on $\widetilde a_i,\widetilde b_i,
\widetilde c_i,\,i\ge1$, is
given in the following theorem.

\begin{thm}\label{Thmofabci}
We have the explicit expressions for  $\widetilde a_m,\, \widetilde b_m,\,
\widetilde c_m,m\ge 1$:
\begin{eqnarray}
\widetilde a_{m+1}&=&\sum_{i=0}^mI_i(<A^{m-i}P_1,BQ_1>-<A^{m-i}P_3,BQ_3>)
-I_{m+1},\ m\ge0,\label{am+1}\\
\widetilde
b_{m+1}&=&\sqrt{2}\sum_{i=0}^mI_i(<A^{m-i}P_1,BQ_2>+<A^{m-i}P_2,BQ_3>)
,\ m\ge0,\label{bm+1}\\
\widetilde
c_{m+1}&=&\sqrt{2}\sum_{i=0}^mI_i(<A^{m-i}P_2,BQ_1>+<A^{m-i}P_3,BQ_2>)
,\ m\ge0,\label{cm+1}\end{eqnarray}
provided that the nonlinearized spatial system (\ref{nxpart}) is satisfied.
Here $I_m,\, m\ge0,$ are defined by
\begin{equation} I_0=1,\
 I_m=\sum _{n=1}^md_n\sum_{\begin{array}{c}{\scriptstyle{
 i_1+\cdots+i_n=m}}\\{ \scriptstyle{i_1,\cdots,i_n\geq
1}}\end{array}}
F_{i_1}\cdots
F_{i_n},\ m\geq1,\label{expressionofI_m}\end{equation}
where the constants $d_n,\,n\ge0,$ are determined recursively by
 \be \left \{
\begin{array}{l}d_1=-\frac 18,\ d_2=\frac3{128},\vspace{1mm}\\
 d_n=-\frac12
\sum_{i=1}^{n-1}d_id_{n-i}-\frac14d_{n-1}-\frac18
\sum_{i=1}^{n-2}d_id_{n-i-1},\
n\ge3,\end{array}\right.\label{dmre}\ee
and the functions $F_m$, $m\ge1$, are given by (\ref{F_m}).
\end{thm}
{\bf Proof}:
By the recursion relation (\ref{recursionre}) and Lemma (\ref{lemmarealre}),
we can obtain that
\begin{eqnarray}&&
 \left(\ba {c} \widetilde c_{m+1}\\ \widetilde b_{m+1}\ea \right)
=\widetilde L^m \left(\ba {c} \widetilde r\\ \widetilde q \ea \right)=
\widetilde L^m \left(\ba {c} \sqrt{2}(<P_2,BQ_1>+<P_3,BQ_2>)\\
\sqrt{2}(<P_1,BQ_2>+<P_2,BQ_3>)\ea \right)\nonumber \\
&&=\sum_{i=0}^mI_i\left(\ba {c} \sqrt{2}(<A^{m-i}P_2,BQ_1>+<A^{m-i}P_3,BQ_2>)\\
\sqrt{2}(<A^{m-i}P_1,BQ_2>+<A^{m-i}P_2,BQ_3>)\ea \right) ,\ m\ge0,\nonumber
\end{eqnarray}
where $I_0=1$ and $I_i,\ 1\le i\le m$, are integrals of motion for
the nonlinearized spatial system (\ref{nxpart}).
Now we compute $\widetilde a_{m+1},\ m\ge0$, by $\widetilde
b_{mx}=-2\widetilde b_{m+1}-2\widetilde q\,\widetilde a_m,\ m\ge0$.
Noting (\ref{nxpart}), we have for $m\ge 0$
 \begin{eqnarray}
\widetilde b_{mx}&=&\sum_{i=0}^{m-1}\sqrt{2}I_i(<A^{m-i-1}P_{1x},BQ_2>+
<A^{m-i-1}P_{1},BQ_{2x}>\nonumber
\\&&\qquad\qquad
+<A^{m-i-1}P_{2x},BQ_3>+<A^{m-i-1}P_{2},BQ_{3x}>)\nonumber\\
&=&\sum_{i=0}^{m-1}\sqrt{2}I_i(<-2A^{m-i}P_1,BQ_2>+<A^{m-i-1}P_1,-\sqrt{2}
\widetilde  qBQ_1>\nonumber\\
&&\qquad\qquad +<\sqrt{2}\widetilde
 qA^{m-i-1}P_3,BQ_3>+<A^{m-i-1}P_2,-2ABQ_3>)\nonumber\\
&=&-2\widetilde b_{m+1}-2\widetilde  q
\Bigl(\sum_{i=0}^{m-1}I_i(<A^{m-i-1}P_1,BQ_1>-<A^{m-i-1}P_3,BQ_3>)
-I_{m}\Bigr),\nonumber
\end{eqnarray}
from which (\ref{am+1}) follows.

In the following we determine the integrals of motion $I_m,\ m\ge0$, by
a relation
\[\widetilde a ^2+
\widetilde b\widetilde c=1,\ \widetilde a=\sum_{i=0}^{\infty}
\widetilde a_i \la ^{-i},\ \widetilde b=\sum_{i=0}^{\infty}
\widetilde b_i \la ^{-i},\ \widetilde c=\sum_{i=0}^{\infty}
\widetilde c_i \la ^{-i},\]
 which gives rise to
\be  2\widetilde a_m= \sum_{i=1}^{m-1}
(\widetilde a_i\widetilde a_{m-i}+\widetilde b_i\widetilde
c_{m-i}),\ m\ge2.\label{usefulre}\ee
First from $\widetilde a_1=0$ we have
\[ I_1=<P_1,BQ_1>-<P_3,BQ_3>=-\frac18 F_1,\]
which shows $d_1=-\frac18$.
Now we suppose $m\ge 2$. At this moment, we have by (\ref{usefulre})
\begin{eqnarray}&&
2\sum_{i=0}^{m-1}I _i\hat a _{m-i}-2I _m
\nonumber \\
&=&\sum_{i=1}^{m-1}
(\sum_{k=0}^{i-1}I _k\hat a_{i-k}-I _i)
(\sum_{l=0}^{m-i-1}I _l\hat a_{m-i-l}-I _{m-i})\nonumber \\
 &&+\sum_{i=1}^{m-1}
\sum_{k=0}^{i-1}I _k\hat b_{i-k}\sum_{l=0}^{m-i-1}I _l
\hat c_{m-i-l}
,\ m\ge2,\nonumber \end{eqnarray}
where $\hat a_{i},\,\hat b_{i},\,\hat c_{i},\,i\ge 1,$ are given by
(\ref{hatabci}).
After interchanging the summing in the above equality, i.e.
\[\sum_{i=1}^{m-1}\,\sum_{k=0}^{i-1}=\sum_{k=0}^{m-2}\,\sum_{i=k+1}^{m-1},\
\sum_{i=1}^{m-1}\,\sum_{l=0}^{m-i-1}=\sum_{l=0}^{m-2}\,\sum_{i=1}^{m-1-l},\
\sum_{i=1}^{m-1}\,\sum_{k=0}^{i-1}\,\sum_{l=0}^{m-i-1}
=\sum_{k=0}^{m-2}\,\sum_{l=0}^{(m-2)-k}\,\sum_{i=k+1}^{m-(l+1)},\]
we may arrive at
\begin{eqnarray}
 -2I_m&= &
\sum_{i=1}^{m-1}I_iI_{m-i}+\sum_{k=0}^{m-2}\sum_{l=0}^{(m-2)-k}
\sum_{i=k+1}^{m-(l+1)}I_kI_{l}
%\times \nonumber\\ &&
(\hat a_{i-k}\hat a_{m-i-l}+\hat b_{i-k}\hat c_{m-i-l})\nonumber\\ &&
-\sum_{k=0}^{m-2}\sum_{i=k+1}^{m-1}I_kI_{m-i}\hat a_{i-k}
-\sum_{l=0}^{m-2}\sum_{i=1}^{m-l-1}I_lI_{i}\hat a_{m-i-l}
-2\sum_{i=0}^{m-1}I_i\hat a_{m-i}\nonumber\\
& :=&B_1+B_2+B_3+B_4+B_5
,\ m\ge2.
\label{delta}\end{eqnarray}
Further we have
\begin{eqnarray}
 B_3&=&-\sum_{k=0}^{m-2}(\sum_{i=k+2}^m +\sum_{i=k+1}-\sum_{i=m})
I_kI_{m-i}\hat a_{i-k}\nonumber\\
&=&-\sum_{k=0}^{m-2}\sum_{i=k+2}^m
I_kI_{m-i}\hat a_{i-k}
-\sum_{k=0}^{m-2}I_kI_{m-k-1}\hat a_1
+\sum_{k=0}^{m-2}I_k\hat a_{m-k}
\nonumber\\&=&-
\sum_{k=0}^{m-2}\sum_{l=0}^{(m-2)-k}I_kI_l\hat a_{m-(k+l)}
-\sum_{k=0}^{m-2}I_kI_{m-k-1}\hat a_1
+\sum_{k=0}^{m-2}I_k
\hat a_{m-k}
,
\nonumber
\end{eqnarray}
\begin{eqnarray} B_4&=&-
\sum_{l=0}^{m-2}(\sum_{i=0}^{m-l-2} -\sum_{i=0}
+\sum_{i=m-l-1})
I_lI_{i}\hat a_{m-i-l}
\nonumber\\
&=&-\sum_{k=0}^{m-2}\sum_{l=0}^{(m-2)-k}I_kI_l\hat a_{m-(k+l)}
+\sum_{l=0}^{m-2}I_l\hat a _{m-l}
-\sum_{l=0}^{m-2}I_lI_{m-l-1}\hat a_1.
\nonumber
\end{eqnarray}
Therefore the latter three terms in the right hand side of
(\ref{delta}) becomes
\[
B_3+B_4+B_5=-2\sum_{k=0}^{m-2}\sum_{l=0}^{(m-2)-k}I_kI_l\hat a _{m-(k+l)}
+\frac14\sum_{\ba {c}{\scriptstyle
k+l=m-1}\\{\scriptstyle  k,l\ge 0 }\ea  }   I_kI_lF_1.
\]
In this way, from (\ref{delta}) we obtain
\begin{eqnarray}
I_m&=&-\frac12 \sum_{i=1}^{m-1}I_iI_{m-i}-\frac18
\sum_{k=0}^{m-2}\sum_{l=0}^{(m-2)-k}I_kI_lF_{m-(k+l)}-\frac18
 \sum_{\ba {c} {\scriptstyle
k+l=m-1}\\{\scriptstyle k,l\ge 0} \ea    } I_kI_lF_1\nonumber\\&
=&-\frac12 \sum_{i=1}^{m-1}I_iI_{m-i}-\frac18
\sum_{\ba {c}{\scriptstyle
k+l\le m-1}\\ {\scriptstyle k,l\ge 0} \ea
  } I_kI_lF_{m-(k+l)},\ m\ge2,\label{star1}
\end{eqnarray}
by which we can determine any $I_m$, $m\ge2$, starting with $I_1=-\frac18F_1$.
It is not difficult to find a homogeneous property among the terms of
 (\ref{star1}). Thus
we may assume that
\[ I_m=\sum _{n=1}^md_n\sum_{\begin{array}{c}{\scriptstyle{
 i_1+\cdots+i_n=m}}\\{ \scriptstyle{i_1,\cdots,i_n\geq
1}}\end{array}}
F_{i_1}\cdots
F_{i_n},\ m\geq2.\]
In general, the coefficients $d_n$, $1\le n\le m$, should depend on $m$.
But the following deduction implies that this assumption is possible.
First  from (\ref{star1}) we easily have
$ I_2=\frac3 {128}F_1^2-\frac18 F_2,$
which leads to  $d_2=\frac 3 {128}$.
When $m\ge 3$, (\ref{star1}) becomes
\be
I_m=-\frac 12\sum_{i=1}^{m-1}I_iI_{m-i}
-\frac18F_m-\frac14\sum_{k=1}^{m-1}I_kF_{m-k}-\frac18 \sum_{\ba {c}
\scriptstyle{k+l\leq m-
1}\\ \scriptstyle{k,l\ge 1}\ea }I_kI_lF_{m-(k+l)},\label{star2}\ee
in which the coefficients of the $F_1^m$ yields the recursion relation
(\ref{dmre}).
In what follows,  we
want to prove that $I_m$, $m\ge 0$, determined above satisfy the relation
(\ref{star2}), indeed.
This may be shown by combining the following three equalities.
First we have
\begin{eqnarray}&&
\sum_{i=1}^{m-1}I_iI_{m-i}=\sum_{i=1}^{m-1}\sum_{k=1}^id_k
\sum_{\ba {c} {\scriptstyle  i_1+\cdots +i_k=i}\\{\scriptstyle  i_1,
\cdots,i_k\ge 1}\ea }F_{i_1}\cdots F_{i_k}
\sum_{l=1}^{m-i}d_l\sum_{\ba {c} {\scriptstyle  j_1+\cdots +j_l=m-i}
\\{\scriptstyle  j_1,
\cdots,j_l\ge 1}\ea }F_{j_1}\cdots F_{j_l}\nonumber\\ &&
=\sum_{k=1}^{m-1}\sum_{i=k}^{m-1}\sum_{l=1}^{m-i}d_kd_l
\sum_{\ba {c} {\scriptstyle  i_1+\cdots +i_k=i}\\{\scriptstyle  i_1,
\cdots,i_k\ge 1}\ea }F_{i_1}\cdots F_{i_k}
\sum_{\ba {c} {\scriptstyle  j_1+\cdots +j_l=m-i}
\\{\scriptstyle  j_1,
\cdots,j_l\ge 1}\ea }F_{j_1}\cdots F_{j_l}\nonumber\\ &&
=\sum_{k=1}^{m-1}\sum_{l=1}^{m-k}\sum_{i=k}^{m-l}d_kd_l
\sum_{\ba {c} {\scriptstyle  i_1+\cdots +i_k=i}\\{\scriptstyle  i_1,
\cdots,i_k\ge 1}\ea }F_{i_1}\cdots F_{i_k}
\sum_{\ba {c} {\scriptstyle  j_1+\cdots +j_l=m-i}
\\{\scriptstyle  j_1,
\cdots,j_l\ge 1}\ea }F_{j_1}\cdots F_{j_l}\nonumber\\ &&
=\sum_{n=2}^{m}\sum_{\ba {c}
 {\scriptstyle  k+l=n}\\{\scriptstyle  k,l\ge 1}\ea }d_kd_l
\sum_{\ba {c} {\scriptstyle  i_1+\cdots +i_n=m}\\{\scriptstyle  i_1,
\cdots,i_n\ge 1}\ea }F_{i_1}\cdots F_{i_n}.
 \nonumber\end{eqnarray}
Similarly we can get the other two equalities
\begin{eqnarray}&&
\sum_{k=1}^{m-1}I_kF_{m-k}=\sum_{n=2}^md_{n-1}
\sum_{\ba {c} {\scriptstyle  i_1+\cdots +i_n=m}\\{\scriptstyle  i_1,
\cdots,i_n\ge 1}\ea }F_{i_1}\cdots F_{i_n}, \nonumber\\&&
\sum_{\ba {c} {\scriptstyle k+l\le m-1}\\{\scriptstyle k,l\ge1
}\ea }I_kI_lF_{m-(k+l)}=
\sum_{n=2}^{m-1}
\sum_{\ba {c} {\scriptstyle i+j=n}\\{\scriptstyle i,j\ge1
}\ea }d_id_j
\sum_{\ba {c}
{\scriptstyle  i_1+\cdots +i_{n+1}=m}\\{\scriptstyle i_1,
\cdots,i_{n+1}\ge 1}\ea }F_{i_1}\cdots F_{i_{n+1}}.
\nonumber\end{eqnarray}
Therefore the proof is finished.
$\vrule width 1mm height 3mm depth 0mm$

\begin{thm}
If $P_i$ and $ Q_i,\ 1\le i\le 3,$ solve
the nonlinearized spatial system (\ref{nxpart}),
then there exist $N$ integrals of motion $\al _i$, $0\le i\le N-1$,
of (\ref{nxpart})
such that
\[
q=\sqrt{2}(<P_1,BQ_2>+<P_2,BQ_3>),\  r=\sqrt{2}(<P_2,BQ_1>+<P_3,BQ_2>)
\]
solve the following $N$-th order stationary AKNS equation
\[
K_N+\sum_{i=0}^{N-1}\alpha _iK_i=0,\]\label{StationaryEquation}
where $K_i,\ 0\le i\le N$, are defined by (\ref{akns}).
\end{thm}
{\bf Proof:} Noting that the expressions (\ref{bm+1}) and (\ref{cm+1}) of
$\widetilde b_{i+1},\,\widetilde c_{i+1},\, i\ge0$,
we can compute that
\begin{eqnarray}
&&\sum_{i=0}^N \al _i \widetilde {K}_i= \sum_{i=0}^N \al _i J
\left ( \begin{array}{c}
\widetilde {c}_{i+1}\\ \widetilde {b}_{i+1} \end{array}\right)
\vspace{2mm}\nonumber
\\
&&=J\sum_{i=0}^N \al _i\sum_{j=0}^i I_j
\left ( \begin{array}{c}
\sqrt{2}(<A^{i-j}P_2,BQ_1>+<A^{i-j}P_3,BQ_2>)\\
 \sqrt{2}(<A^{i-j}P_1,BQ_2>+<A^{i-j}P_2,BQ_3>)\end{array}\right) \nonumber
\vspace{2mm}
\\
&&=\sqrt{2}J\sum_{i=0}^N \al _i\sum_{j=0}^i I_j\sum_{k=1}^N\la _k^{i-j}
\mu _k
\left ( \begin{array}{c}
\phi _{2k}\psi _{1k}+\phi _{3k}\psi _{2k}
\\
\phi _{1k}\psi _{2k}+\phi _{2k}\psi _{3k}
\end{array}\right)\nonumber
\vspace{ 2mm}\\
&&
=\sqrt{2}J\sum_{k=1}^N\mu _k
\Bigl(\sum_{i=0}^N \sum_{j=0}^i \alpha _i I_j\la _k^{i-j}
\Bigr)\left ( \begin{array}{c}
\phi _{2k}\psi _{1k}+\phi _{3k}\psi _{2k}\\
\phi _{1k}\psi _{2k}+\phi _{2k}\psi _{3k}
\end{array}\right)\nonumber
\vspace{ 2mm}
\\
&&=\sqrt{2}J\sum_{k=1}^N\mu _k\sum _{l=0}^N\Bigl(
\sum_{i=l}^N \al _i I_{i-l}\Bigr)\la _k^{l}
\left ( \begin{array}{c}
\phi _{2k}\psi _{1k}+\phi _{3k}\psi _{2k}\\
\phi _{1k}\psi _{2k}+\phi _{2k}\psi _{3k}
\end{array}\right).\nonumber
\end{eqnarray}
Secondly we set
\[
G(\la )=\prod _{i=1}^N (\la -\la _i)
=\sum _{i=0}^N \beta _i \la ^i=\la ^N+\sum _{i=0}^{N-1}\beta _i \la ^i.
\]
Let us now choose
\[
\sum_{i=l}^N\al _i I_{i-l}=\beta _l,\ 0\le l \le N,\]
which determines recursively
\[
\al _N=\beta _N=1,\ \al _l=\beta _l-\sum _{i=l+1}^N\al _iI_{i-l},\ 0\le l\le
N-1,\]
due to $I_0=1$. The $\al _i,\ 0\le i\le N-1,$ are all integrals of motion
of (\ref{nxpart}) since they are functions of $I_i$,
$0\le i\le N$.
Further by $G(\la _k)=0,\ 1\le k\le N$, we see that $\sum_{k=0}^N\al _i
\widetilde K _i=0$, which
completes the proof.
$\vrule width 1mm height 3mm depth 0mm$

The above theorem also implies that the potential determined by the
Bargmann symmetry constraint (\ref{Bargmannsym})
is a finite gap potential of the
spectral problem (\ref{newsp}).

\begin{thm}
Under the control of the nonlinearized spatial system (\ref{nxpart})
,
the nonlinearized temporal systems (\ref{ntpartn})
for $n\ge 0$ can also be
rewritten as the Hamiltonian systems
\be  P  _{it_n}=\{P_i,H_n\}=-B^{-1}
\frac {\part H_{n}}{\part  Q  _i},\
 Q  _{it_n}=\{Q_i,H_n\}=B^{-1}\frac {\part H_{n}}{\part P  _i},\
\ i=1,2,3\label{hamilofntpartn}\ee
with the Hamiltonian functions
\[H_n= -\frac14
\sum _{m=0}^n \frac {d_m}{m+1}
\sum_{\begin{array}{c}{\scriptstyle{ i_1+\cdots+i_{m+1}=n+1}}
\vspace{-1mm}
\\ {\scriptstyle{ i_1,\cdots,i_{m+1}\ge1}}\end{array}}
F_{i_1}
\cdots F_{i_{m+1}}\, ,\]
where $d_0=1$ and $F_m,\,m\ge1$, are defined by (\ref{F_m}).
\end{thm}
{\bf Proof}: We only prove the former equality of (\ref{hamilofntpartn}).
 We know that
under the control of the nonlinearized spatial system (\ref{nxpart}),
the results in Theorem \ref{Thmofabci} holds. Hence
we have
\begin{eqnarray}
P_{1t_n}&=&2\sum_{i=0}^n\widetilde a_iA^{n-i}P_1+\sqrt{2}\sum_{i=0}^n
\widetilde b_iA^{n-i}P_2\nonumber\\
&=&-2A^nP_1+2\sum_{i=1}^n
(\sum_{k=0}^{i-1}I _k\hat a_{i-k}
-I _i)
A^{n-i}P _1
 +\sqrt{2}\sum
_{i=1}^n\sum_{k=0}^{i-1}I _k \hat b_{i-k}
A^{n-i} P_2\nonumber\\&
=&-2\sum _{k=0}^nI _kA^{n-k}P_1
+\sum_{k=0}^{n-1}I _k\sum_{i=k+1}^n2\hat a_{i-k}
A^{n-i}P_1
+\sum_{k=0}^{n-1}I _k\sum_{i=k+1}^n
\sqrt{2}\hat b_{i-k}
A^{n-i}P_2\nonumber\\&
=&\frac14 I _nB^{-1}\frac {\part F_1}{\part Q_1}+\frac14
\sum_{k=0}^{n-1}I _kB^{-1}\frac {\part
F_{n-k+1}}{\part Q_1}=\frac14
\sum_{k=0}^{n}I _kB^{-1}\frac {\part F_{n-k+1}}{\part Q_1},
\nonumber\end{eqnarray}
where $\hat a_i,\,\hat b_i,\,i\ge1,$ are given by (\ref{hatabci}).
We further note the expression
of $I_m,\ m\ge0$, defined by (\ref{expressionofI_m})
and then  we may make the following performance
\begin{eqnarray}
P_{1t_n}&=&\frac14I _0B^{-1}\frac {\part F_{n+1}}{\part Q_1}+\frac14
\sum _{k=1}^n\sum
_{m=1}^kd_m\sum_{\ba {c}{\scriptstyle i_1+\cdots+i_m=k}\\ {\scriptstyle
i_1,\cdots,i_m\ge1}\ea }
F_{i_1}
\cdots F_{i_m}B^{-1}\frac {\part F_{n-k+1}}{\part Q_1}\nonumber\\&
=& \frac14I _0B^{-1}
\frac {\part F_{n+1}}{\part Q_1}+\frac14\sum _{m=1}^n d_m\sum
_{k=m}^n\,\sum_{\ba {c} {\scriptstyle i_1+\cdots+i_m=k}\\ {\scriptstyle
i_1,\cdots, i_m \ge1}\ea }
F_{i_1}
\cdots F_{i_m}B^{-1}\frac {\part F_{n-k+1}}{\part Q_1}\nonumber\\&
=& \frac14I _0B^{-1}
\frac {\part F_{n+1}}{\part Q_1}+\frac14\sum _{m=1}^n \frac
{d_m}{m+1}B^{-1}\frac \part {\part Q_1}
\sum_{\ba {c}{\scriptstyle i_1+\cdots+i_{m+1}=n+1}\\ {\scriptstyle
 i_1,\cdots,i_{m+1}\ge1}\ea }
F_{i_1}
\cdots F_{i_{m+1}}\nonumber\\&
=&\frac14B^{-1}\frac \part {\part Q_1}
\sum _{m=0}^n \frac {d_m}{m+1}
\sum_{\ba {c}{\scriptstyle  i_1+\cdots+i_{m+1}=n+1}\\ {\scriptstyle
i_1,\cdots,i_{m+1}\ge1}\ea }
F_{i_1}
\cdots F_{i_{m+1}}
=-B^{-1}\frac {\part H_n}{\part Q_1},
\nonumber\end{eqnarray}
where we have accepted $d_0=1$. The above manipulation is fulfilled
for the case of $n\ge 1$. The case of $n=1$ needs only a simple calculation.
Thus the former equality of (\ref{hamilofntpartn}) is true for $n\ge 0$.
The latter equality of (\ref{hamilofntpartn}) may be proved similarly.
The proof is completed.
$\vrule width 1mm height 3mm depth 0mm$

The above theorem  allows us to establish  a sort of  involutive
 solutions to AKNS soliton equations, which exhibits a kind
of separation of variables for AKNS soliton equations. This
is the following result.

\begin{thm}
The $n$-th AKNS soliton equation $u_{t_n}=K_n$ has the involutive solution
with separated variables $x, t_n$
\be  \left \{ \begin{array}{l}
q=\sqrt{2}(<g^x_{H}g^{t_n}_{H_n}P_1(0,0),
Bg^x_{H}g^{t_n}_{H_n}Q_2(0,0)>\vspace{2mm}\\
\qquad\qquad+<g^x_{H}g^{t_n}_{H_n}P_2(0,0),
Bg^x_{H}g^{t_n}_{H_n}Q_3(0,0)>),\vspace{3mm}
\\
r=\sqrt{2}(<g^x_{H}g^{t_n}_{H_n}P_2(0,0),
Bg^x_{H}g^{t_n}_{H_n}Q_1(0,0)>\vspace{2mm}
\\ \qquad\qquad+<g^x_{H}g^{t_n}_{H_n}P_3(0,0),
Bg^x_{H}g^{t_n}_{H_n}Q_2(0,0)>).\end{array}\right.\label{involutivesolu}\ee
where $g^y_G$ denotes the Hamiltonian phase flow of $G$ with a variable
$y$ and $P_i(0,0)$ and $Q_i(0,0),\ 1\le i\le 3$,
 may be arbitrary initial value vectors.
\end{thm}
{\bf Proof}:
Let
\[ P_i(x,t_n)=g_H^xg^{t_n}_{H_n}P_i(0,0),\ Q_i(x,t_n)=
g_H^xg^{t_n}_{H_n}Q_i(0,0),\ 1\le i\le 3.\]
Then $P_i(x,t_n)$ and $Q_i(x,t_n),\ 1\le i\le 3$, solve the nonlinearized
spatial system (\ref{nxpart}) and the Hamiltonian system
(\ref{hamilofntpartn}). However under the control of (\ref{nxpart}),
(\ref{hamilofntpartn}) is equivalent to the nonlinearized temporal system
(\ref{ntpartn}). This shows that $P_i(x,t_n)$ and $Q_i(x,t_n)$ also solve
(\ref{nxpart}) and (\ref{ntpartn}), simultaneously. Therefore
 the compatibility condition of (\ref{nxpart}) and (\ref{ntpartn})
is satisfied, i.e. (\ref{involutivesolu}) determines a solution to
$u_{t_n}=K_n$. In addition, since $\{H,H_n\}=0$, the Hamiltonian phase
flows $g^x_H,\,g^{t_n}_{H_n}$ may commute with each other.
It follows that  the resulting solution (\ref{involutivesolu}) is involutive.
The proof is finished.
$\vrule width 1mm height 3mm depth 0mm$

\section{Conclusions and remarks}
\setcounter{equation}{0}

We have introduced a three-by-three matrix  spectral problem for the usual
AKNS soliton hierarchy and proposed the corresponding
 Bargmann symmetry constraint on this AKNS hierarchy.
Moreover we have
exhibited an explicit Poisson algebra
\be \{ \bar F_j,\,1\le j\le N,\,
F_m,\,m\ge1\}\label{Poissonalgebra}\ee
 on the
symplectic manifold $(\R^{6N},\omega ^2)$ and further
a binary nonlinearization procedure is manipulated along with
 a sort of  involutive
 solutions to AKNS soliton equations.
When $N=1,2$, we have proved the nonlinearized spatial system (\ref{nxpart})
is integrable in the Liouville sense, indeed.
But we don't know if we can take out enough independent integrals of motion
among the Poisson algebra (\ref{Poissonalgebra})
 for a general integer $N$.
We hope that this Poisson algebra suffices for proving complete integrability
of the nonlinearized Lax systems.

It should be pointed out that
the Neumann symmetry constraint
and the
higher order symmetry constraints
\be K_{-1}=J\sum _{j=1}^NE_j\frac {\delta \la _j}{\delta u},\
K_m=JG_m=J\sum _{j=1}^NE_j\frac {\delta \la _j}{\delta u},\ (m\ge1),\ee
may also be considered.
These sorts  of symmetry constraints are somewhat different from the
Bargmann symmetry constraints because $K_{-1}$ is a constant vector
and the conserved covariants
$G_m,\, m\ge1, $ involve the differential of the potential $u$ with
respect to the space variable $x$.
In order to discuss them, we are required to introduce
a new symplectic submanifold of the Euclidean
spaces in the case of the Neumann constraint
and
new dependent variables,
i.e. the so-called Jacobi-Ostrogradsky coordinates \cite{Deleon},
in the case of higher order constraints.
Similarly, we can consider the corresponding
 $\tau$-symmetry (time first order dependent
symmetry) constraints or more generally,
time polynomial dependent symmetry constraints.
Note that the similar Bargmann
symmetry constraints have also been carefully analyzed
for KP hierarchy \cite{OevelStrampp}
and the
symmetries in the right hand side of the Bargmann  symmetry constraints
are sometimes called additional symmetries \cite{Dickey}
 and may be taken as
sources of soliton equations
\cite{Mel'nikov}.

We remark that the finite dimensional Hamiltonian systems
generated by nonlinearization technique depend
on the starting spectral
problems. Therefore the same soliton equation may relate to different
finite dimensional Hamiltonian systems once it possesses different
Lax representations. AKNS soliton equations are exactly such examples.
But we don't know
if there exists an interrelation among
the different finite dimensional Hamiltonian systems generated from
the same soliton equation.
In the binary nonlinearization procedure itself, there also exist some
intriguing open problems. For example,
why do the nonlinearized spatial system
and the nonlinearized temporal systems for $n\ge 0$
under the control of the nonlinearized
spatial system always possess Hamiltonian structures? We don't know either
whether or not the nonlinearized temporal systems
for $n\ge 0$
are themselves
integrable soliton equations
without the control of the nonlinearized spatial system.
These problems are worth studying in order to enrich integrable
structures of soliton equations.

\noindent{\bf Acknowledgments:} One of the authors (W. X. Ma)
would like to thank the
Alexander von Humboldt Foundation for a research fellow award and
the National Natural
Science Foundation of China and the Shanghai Science and Technology Commission
of China
 for financial support.
He is also grateful to Drs. P. Zimmermann
and G. Oevel for their helpful and stimulating discussions about
MuPAD.


\begin{thebibliography}{99}
\bibitem{AdamsHP} M. R. Adams, J. Harnad and E. Previato,
Isospectral Hamiltonian flows in finite and infinite dimensions,
 I. Generalized Moser systems and moment maps into loop algebras,
Commun. Math. Phys.
117 (1988) 451--500.
\bibitem{AntonowiczWojciechowski1} M. Antonowicz and S. Rauch Wojciechowski,
Restricted flows of soliton hierarchies: coupled KdV and Harry Dym case,
 J. Phys. A: Math. Gen.  24 (1991) 5043--5061(1991).
\bibitem{AntonowiczWojciechowski2} M. Antonowicz and S. Rauch Wojciechowski,
How to construct finite dimensional bi-Hamiltonian systems from soliton
equations: Jacobi integrable potentials, J. Math. Phys. 33 (1992) 2115--2125.
\bibitem{Arnold} V. I. Arnold, Mathematical Methods in Classical Mechanics
(Springer-Verlag, New York, 1980).
\bibitem{BogoNovikov} O. I. Bogoyavlenskii and S. P. Novikov,
The relationship between Hamiltonian formalism of stationary and non
stationary
problems, Funct. Anal. Appl. 176 (1976) 8--11.
\bibitem{Blaszak} M. Blaszak, Bi-Hamiltonian field Garnier system,
  Phys. Lett. A  174 (1993) 85--88;
Miura map and bi-Hamiltonian formulation for restricted flows
 of the KdV hierarchy, J. Phys. A: Math. Gen.  26 (1993) 5985--5996.
\bibitem{Cao} C. W. Cao,
Nonlinearization of the Lax system for AKNS hierarchy, Sci. China A 33 (1990)
 528--536.
\bibitem{CaoGeng} C. W. Cao and X. G. Geng, Classical integrable systems
generated through nonlinearization of eigenvalue
problems, in: Nonlinear Physics, eds. C. H. Gu, Y. S. Li and G. Z. Tu,
(Springer Verlag, Berlin, 1990) 68--78.
\bibitem{Carroll} R. W. Carroll, Topics in Soliton Theory (North-Holland,
Amsterdam, 1991).
\bibitem{Dickey} L. A. Dickey, On the constrained KP hierarchy,
Lett. Math. Phys. 34 (1995) 379--384.
\bibitem{Flaschka} H. Flaschka, Toward an algebro-geometrical
interpretation of the Neumann system, Tohoku Math. J. 36 (1984) 407--426.
\bibitem{FokasAnderson} A. S. Fokas and R. L. Anderson, On the use of
isospectral eigenvalue problems for obtaining hereditary symmetries
for Hamiltonian systems,
J. Math. Phys. 23
(1982) 1066--1073.
\bibitem{FuchssteinerFokas} B. Fuchssteiner and A. S. Fokas,
Symplectic structures, their B\"acklund
transformations and hereditary symmetries, Physica D 4 (1981) 47--66.
\bibitem{Geng} X. G. Geng, The Hamiltonian structure
and new finite-dimensional integrable system associated
with Harry-Dym type equations, Phys. Lett. A 194 (1994) 44--48.
\bibitem{GengMa} X. G. Geng and W. X. Ma, A generalized Kaup-Newell spectral
problem, soliton equations and finite-dimensional integrable systems,
Nuovo Cimento A 108 (1995) 477--486.
\bibitem{Deleon} M. de Le\'on and P. R. Rodrigues, Generalized Classical
Mechanics and Field Theory (North-Holland, Amsterdam, 1985).
\bibitem{Ma} W. X. Ma,
New finite dimensional integrable systems by symmetry constraint
of the KdV equations, J. Phys. Soc. Jpn. 64 (1995) 1085--1091;
Symmetry constraint of MKdV equations by binary nonlinearization,
to appear in Physica A.
\bibitem{MaStrampp} W. X. Ma and W. Strampp,
An explicit symmetry constraint for Lax pairs
and the adjoint Lax pairs of AKNS systems, Phys. Lett. A 185 (1994)
277--286.
\bibitem{Mel'nikov} V. K. Mel'nikov, New method for deriving nonlinear
integrable systems, J. Math. Phys. 31 (1990) 1106--1113.
\bibitem{OevelStrampp} W. Oevel and W. Strampp, Constrained KP hierarchy and
bi-Hamiltonian structures, Commun. Math. Phys. 157 (1993) 51--81.
\bibitem{Perelomov} A. M. Perelomov, Integrable Systems of
Classical Mechanics and Lie Algebras, Vol. I, (Birkhaeuser Verlag, Basel,
1990).
\bibitem{RagnicoWojciechowski} O. Ragnisco and S. Rauch-Wojciechowski,
Restricted flows of AKNS hierarchy,  Inverse Problems  8 (1992) 245--262.
\bibitem{Schilling} R. J. Schilling, Generalizations of the Neumann system,
a curve-theoretical approach--part I, Commun. Pure Appl. Math. 40 (1987)
455--522.
%\bibitem{Schilling2} R. J. Schilling,
%Neumann system for the algebraic AKNS problem,
\bibitem{Tondo} G. Tondo, On the integrability of finite-dimensional systems
related to soliton equations (preprint, Quaderno n. 340, Aprile 1995).
\bibitem{Tu} G. Z. Tu,
The trace identity, a powerful tool for constructing the Hamiltonian
structure of integrable systems,
J. Math. Phys. 30 (1989) 330--338.
\bibitem{TuAH} G. Z. Tu, R. I. Andrushkiw and X. C. Huang, A trace identity
and its application to integrable systems of $1+2$ dimensions,
J. Math. Phys. 32 (1991) 1900--1907.
\bibitem{Zeng1} Y. B. Zeng, An approach to the deduction of the
finite-dimensional integrability from the infinite-dimensional integrability,
Phys. Lett. A 160 (1991) 541--547.
\bibitem{Zeng2} Y. B. Zeng, New factorization of the Kaup-Newell hierarchy,
Physica D 73 (1994) 171--188.

\end{thebibliography}
\end{document}